\documentclass[conference]{IEEEtran}
\usepackage{stmaryrd}
\usepackage{amsfonts}
\usepackage{array}

\usepackage{graphicx,times,psfig,amsmath} 

\IEEEoverridecommandlockouts    

\begin{document}

\title{\ \\ \LARGE\bf Augmented Neural Networks for Modelling Consumer Indebtness
\thanks{Alexandros Ladas, Jon Garibaldi, Rodrigo Scarpel and Uwe Aickelin are with the School of Computer Science, University of Nottingham, Jubilee Campus, Wollaton Road, Nottingham, NG8 1BB, UK (email: \{psxal2, uwe.aickelin\}@nottingham.ac.uk,jmg@cs.nott.ac.uk, rodrigo@ita.br)} }

\author{Alexandros Ladas, Jon Garibaldi, Rodrigo Scarpel and Uwe Aickelin}


\maketitle

\begin{abstract}
Consumer Debt has risen to be an important problem of modern societies, generating a lot of research in order to understand the nature of consumer indebtness, which so far its modelling has been carried out by statistical models. In this work we show that Computational Intelligence can offer a more holistic approach that is more suitable for the complex relationships an indebtness dataset has and Linear Regression cannot uncover. In particular, as our results show, Neural Networks achieve the best performance in modelling consumer indebtness, especially when they manage to incorporate the significant and experimentally verified results of the Data Mining process in the model, exploiting the flexibility Neural Networks offer in designing their topology. This novel method forms an elaborate framework to model Consumer indebtness that can be extended to any other real world application.
\end{abstract}


\section{Introduction}

\PARstart{C}{onsumer} Debt Analysis has received recently a lot of attention from the research community 
in an effort to explain the ``nature'' of consumer indebtness that has emerged recently in the developed countries. Among the three fundamental research questions posed in the analysis of this social problem \cite{livin92} lies the identification of factors that affect the level of consumer debt. Answering the latter, ongoing research revealed a series of diverse factors, economic, demographic and psychological, that are related to how deep a consumers goes in debt \cite{kam12,stone06,wang11,kam07} providing a deep insight in the ``nature'' of this problem.

The discovery of these factors was mainly carried out by traditional statistical models like linear regression which has the ability to reveal linear associations between variables. However, as common as the utilisation of these models in the field of Economics might be, so is their limited ability to deal with characteristics that data from real world applications possess. Their difficulty to handle non-linearity in the data makes them unable to solve non-linear classification problems \cite{ref94}, while the colinearity between the independent variables can lead to incorrect identifications of most predictors \cite{sousa07}. These limitations  make them inappropriate to model successfully consumer indebtness since socio-economic datasets exhibit strong non-linearity among several other inconsistencies. It also raises questions regarding the validity of the relationships uncovered by these models as their small predictive accuracy cannot guarantee the identification of the correct predictors. In addition to this, most of the research has been conducted on a limited number of observations making hard to consider the findings as representative.

As the need to develop fairly accurate quantitative prediction models becomes apparent \cite{ati01}, we argue that the field of Economics can benefit from the variety of techniques and models Computational Intelligence has to offer. Such a computational model is the Neural Networks, a system of interconnected ``neurons'', inspired by the functioning of the central nervous system. Neural networks are capable of machine learning and not only they manage to achieve remarkable prediction accuracy by successfully handling non-linearity in the data but their flexibility in the design of their topology also offers a way to incorporate important steps of the Data Mining process into a regression model. The potential of Data Mining is evident in the numerous ways to pre-process the data in order to tackle any inconsistencies they may contain and to explore the relationships in the data, that be can combined in an elaborate process for Knowledge Discovery in any difficult real world problem like consumer indebtness. 

Therefore in order to evaluate the impact Neural Networks can make on modelling the Consumer Debt in a large socio-economic dataset  in this work, we compare their performance against Random Forests and linear regression. In the same experimental setup we also evaluate the contribution on the performance of these models of a series of Data Mining techniques like the transformations performed on the data in order to deal with the inconsistencies they contain, such noise, high dimensionality and the presence of outliers and the a classification of debtors identified by clustering. Finally we take advantage of the ability to design the topology of Neural Networks and we introduce a novel way to incorporate into the topology meaningful information that derives from explanatory techniques applied on data, like Clustering and Factor Analysis, and we assess its performance. 

Our results show that the transformations on the data improve in a great extend the accuracy of all three regression models and that Neural Networks achieve the best performance. The contribution of the classifications provided by clustering remains argumentative when it is used as an extra variable but proves to be very useful when it is incorporated in an appropriate way in the topology of the Neural Networks which leads to a further improvement in the performance of the model. Therefore, we believe that this work not only serves as a comparison between Neural Networks and other regression models but it also verifies the great of potential of Neural Networks that can be strong predictors and take advantage of significant results from Data Mining methods at the same time, sketching a complete framework for the Consumer Debt Analysis including necessary transformations of data, exploratory models and reliable regression model that it may extend to any real world application problem that contains a dataset with similar inconsistencies and characteristics as this one.

The rest of the paper is organised as following. In the 2nd section we discuss the related work on the level of debt predictions and on the models we use for our purposes. In the 3rd section we introduce briefly the CCCS dataset together with transformations performed on its attributes and the clustering approach that identified classes of debtors. We then present the models in the 4th section whereas in the 5th we proceed with the details of the experimental set up. Finally in the 6th section we analyse the results of our experiments and we conclude our work in the 7th section.

\section{Related Work}
Statistical models and linear regression are primarily used for the level of debt prediction in the literature. A significant amount of the work is summarised in \cite{stone06} where they also provide a model for separating debtors from non-debtors. However, their suggested logit model suffers from a low $R^2$ (33\%). In a similar way, in \cite{gather12,ott11} the proposed models that take into account psychological factors as  predictors,  exhibit even lower $R^2$ in their probit models (around 10\%). Surprisingly enough the linear regression model presented in \cite{livin92} achieves a remarkable 66\% $R^2$ but as it is explained in \cite{stone06}, this big proportion of variance explained, is due to the small number of respondents. A linear regression model built for estimating the outstanding credit card balance in \cite{kim01} exhibits 30\% $R^2$. Based on these results and the fact that the models are built on a limited number of observations, we are unsure whether to regard these findings as reliable since the suggested models fail to explain the variance that exists in the data and the small number of instances cannot be considered representative enough. This is further enhanced by the criticism statistical techniques receive in \cite{ref94}, where it is argued that they have reached their limitations in applications with datasets that contain non-linearity in the data, like an indebtness dataset.

On the other hand, Random Forests, a popular machine learning algorithm  for Data Mining, has been shown to be able to handle non-linearities in the data \cite{grom09}. They have received a lot of attention in biostatistics and other fields \cite{grom09} due to their ability to handle a large number of variables with a relatively small number of observations and because they provide a way to identify variable importance \cite{grom09,seg04}. They manage to demonstrate exceptional performance with only one parameter and their regression has been proven not to overfit the data \cite{seg04}. An interesting application of Random Forests is in \cite{ghos11} where a model measuring the impact of the reviews of products in sales and perceived usefulness was constructed.

Similarly, Neural Networks exhibit better generalisation than linear regression models \cite{ref94,sousa07},  allow for extrapolation \cite{sousa07} and can handle non-linearity \cite{ref94} posing as strong predictors.  Their huge learning capacity has led many of researchers to believe that they are able to approximate any function that is encountered in applications \cite{horn89,ding10}. They have been shown to outperform Linear Regression models \cite{ref94,sousa07} and in  Economics they have been successfully used for stock performance modelling \cite{ref94} and for credit risk assessment \cite{ati01}. A very interesting ability they possess is the ability to fully parametrise the topology of the network introducing a concept of logical structure among the neurons that consist the network. This has been exploited in \cite{ding10} where Factor Analysis is utilised in order to define the topology of the network and although their result has shown not to actually improve the precision of the existing neural network, it manages to speed up the convergence of the algorithm. The same idea has been adopted by us in this work for further experimentation in our dataset and has been extended in order to include further information that derives from clustering the data. As Neural Networks have not been used so far for the purposes of Consumer Debt Analysis, in this work we exploit the many advantages they offer in order to achieve a better modelling of consumer indebtness than the existing ones, supporting their utilisation in the field of Economics, in applications of which they already have replaced traditional econometric models.






.



\section{CCCS Dataset}
\subsection{Description}
The CCCS dataset, introduced in \cite{dis09}, is a socioeconomic crossectional dataset based on the data provided by the Consumer Credit Counseling Service. Its 58 attributes contain information about approximately 70000 clients who contacted the service between the years 2004 and 2008 in order to require advice about how they can overcome their debts. The information was gathered through interviews when each client first contacted the service and it varies from standard
demographics to financial details, aggregated spending in categories and debt details. The attributes of interest for the purpose of Consumer Debt Analysis are limited to Demographics, Expenditure and Financial attributes as they can be seen in Table I together with their description.

\begin{table}[htbp]
\caption{Description of CCCS attributes}
\begin{center}
\begin{tabular}{ll}
\hline\noalign{\smallskip}
$ Attribute $ & $ Description $ \\
\noalign{\smallskip}
\hline
\noalign{\smallskip}

pid & individual identifier\\
{\bfseries Demographics} \\
age & age of person\\
mstat & marital status\\
empstat & employment status\\
male & sex of person\\
hstatus & housing status \\
ndep & number of dependants in household\\
nadults & number of adults in household\\
{\bfseries Financial Attributes}\\
udebt & total value of unsecured debt\\
mortdebt & total value of mortgage debt\\
hvalue & total value all housing owned \\
finasset & total value of financial assets\\
carvalue & resale value of car\\
income & total monthly income\\
{\bfseries Expenditure} \\
clothing & total monthly spending on clothing\\
travel & total monthly spending on travel\\
food & total monthly spending on food\\
services & total monthly spending on utilities\\
housing & total monthly spending on housing\\
motoring & total monthly spending on motoring\\
leisure & total monthly spending on leisure\\
priority & total monthly spending on priority debt\\
sundries & total monthly spending on sundries\\
sempspend & total monthly self-employed spending\\
other & total other spending\\
{\bfseries Debt Details}\\
ndebtitems & number of debt items\\

\hline
\end{tabular}
\end{center}
\end{table}

\subsection{Transformations}
Like other real world dataset, CCCS contains noise and outliers, while at the same time it suffers from high dimensionality. In order to tackle the aforementioned difficulties a series of transformations steps were performed in an earlier work \cite{lad14} that proved to be beneficial for the unsupervised approach of this dataset.  More precisely, Homogeneity analysis (Homals) \cite{lee09} was utilised in order to map the categorical demographic data, significant attributes concerning the Consumer Debt Analysis, into two-dimensional coordinates together with a Factor analysis on the financial attributes and a clustering on the correlation of the spending items. These transformations reduced the dimensionality to more compact attributes, removed noise and outliers, provided a sense of  interpretability and improved the quality of the clustering. A summary of the transformations can be seen in Fig.1 whereas the new nine transformed attributes include two spatial coordinates that discriminate the Demographic variables, three Financial Factors that summarise all the informations that lies in Financial Attributes and four Behavioural Spending Clusters that characterise spending in Necessity, Household, Excessive and Leisure.

\begin{figure}[btp]
\centerline{\includegraphics[width=3.0in]{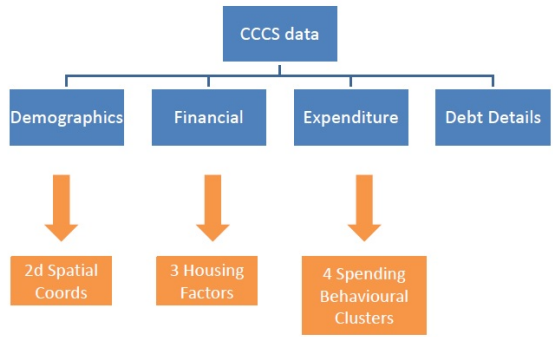}}
\caption{Transformations of CCCS attributes} \label{fig 1}
\end{figure}

\subsection{Classification of Debtors}
Finally, in \cite{lad14} these transformations were proved to be useful for  the clustering of a random sample of 10000 debtors from the CCCS dataset that managed to classify 8370 debtors in seven classes with distinct characteristics. The characteristics of these classes can be seen in the Table II, which also includes the 1630 debtors that remained unclassified. Further information regarding the dataset itself, the suggested transformations and the clustering results can be found in \cite{lad14} as it is not the subject of this work. Our objective is to use the information that derives from the exploratory research that was conducted in \cite{lad14}, meaning the transformed attributes and classifications, in order to evaluate their contribution in the level of debt prediction.

\begin{table}[tpb]
 
  \caption{Description of classes of Debtors}
	\begin{center}  
  \begin{tabular}{ll<{\raggedright\arraybackslash}p{3cm}}
    \hline\noalign{\smallskip}
    $ Class $ & $ Size $  & $ Characterisation $ \\
    \noalign{\smallskip}
    \hline
    \noalign{\smallskip}
    1     & 2301  & Young single unemployed debtors with low income, debt and spending \\
    2     & 1440  & Average Income-spending- debt debtors usually p/t employed and cohabiting with high spending in clothing and food\\
    3     & 1033  &  High Income-Debt-Spending Debtors, usually self-employed and with expensive houses\\
    4     & 948   & Older and retired debtors with average income-spending and low levels of debt \\
    5     & 507   & High Income-Debt-Spending Debtors with cheap houses\\
    6     & 1588  & Average Income-spending-debt debtors usually p/t employed but single, divorced or separated\\
    7     & 553   & Old and retired Debtors with low income, debt and spending, other marital status \\
    8     & 1630  & Unclassified\\
    \hline
    \end{tabular}%
  \end{center}
\end{table}%


\section{Models}
\subsection{Linear Regression}
Linear Regression is the simplest of the statistical models and it tries to model the relationship between a dependant variable and one or more explanatory variables. As someone can refer from the name, Linear Regression assumes a linear relationship between the dependant variable and the explanatory variables and tries to fit a straight line in the data. More formally Linear Regression is defined as:

\begin{equation}
Y= \beta_{0} + X_{1}\beta_{1} +....+ X_{p}\beta_{p} + \epsilon,
\end{equation}

\noindent where $\beta_{0}$, $\beta_{1}$, ...., $\beta_{p}$ are the coefficients and  $X_{j}$,j=1,....p denote p regressor variables. Finally $ \epsilon $ denotes the error term which is assumed to be uncorrelated to the regressors and have mean and variance equal to 0. The model takes as input the observations and tries to fit the straight line by estimating the parameters (coefficients and error term). A widely used algorithm for estimating the parameters is the Ordinary Least Squares(OLS) which tries to minimise the sum of squared residuals.

\subsection{Random Forest Regression}
Random Forest is an example of ensemble learning that generates many classifiers and aggregate the results \cite{bre01}. The Random Forest method creates large number of Decision Trees for the case of classification or Regression Trees for the case of regression from different random samples of the data. The samples are being drawn based on bootstrap techniques that allow resampling of instances. The appropriate tree is being constructed based on each sample and its accuracy is evaluated on the rest of the samples. The difference from the common Decision Tree is that when a split on a node is to be decided, a specific number of the attributes can participate as candidates and not all of them. When the random forest is built the prediction is made by aggregating the votes of all the trees for the case of classification and by averaging the results of all the trees for the case of regression. It needs the specification of only two parameters, the size of the forest and the number of predictors that can be candidates for each node split and its success is based on its simplicity. The notion of randomness it adopts in its process allows the model to be robust against data overfitting.

\subsection{Neural Networks}
A Neural Network is a directed graph consisting of nodes and edges that are organised in layers. As it models a relationship between the predictors and the response variables, the input layer is consisted of nodes that represent the predictors and the output layer of nodes that represent the response variables if there are more than one. One or more hidden layers of an arbitrary number of nodes  connect these two layers. Each layer is fully connected with the next layer and each edge assigns a weight to the value it takes as input and passes it on the next node. Thus in each node the weighted sum of all the nodes that belong to the previous layer is calculated adding the intercept and the result is being fed into an activation function and passed to the next layer. The activation function is usually a non-linear activation function like the sigmoid function or the hyperbolic tangent. The simplest Neural Network (Perceptron) has n inputs and one output and it is identical to the logistic regression as it is a non-linear function of the linear aggregation of the input. With this in mind we can easily conclude that a Neural Network with more than one node in the hidden layer is an extension of the Generalised Linear Models.  



A Neural Network takes as parameters the starting weights of the edges that are usually initialised randomly and the network topology meaning the organisation of the nodes in the hidden layers. Then the model tries to find the optimal weights of the edges by using a learning algorithm like Backpropagation on the data. Backpropagation tries to minimise the difference between the predicted value calculated by the model and the actual value. It does that by calculating this difference and then following the chain rule it moves from the output to the input adapting all the appropriate weights according to a specific learning rate. Resilient Backpropagation which is argued to be more suitable for regression purposes \cite{gun10} is similar to Backpropagation but instead of subtracting a ratio of the gradient of the error function like Backpropagation does, it increases the weight if the gradient is negative and reduces it if its positive.  It updates the weights by using only the sign of the gradient and some predefined values. The value of the update is bigger if the gradient changes sign from the previous update and smaller if it keeps the same sign. This way it ensures that a local minimum won't be missed.

The Neural Networks tend to overfit the data, a fact that raises a concern of how they can be properly used. A common technique for avoiding data overfitting is to train the model on a subset of the data and validate it on the rest of the data. A very popular technigue in Supervised Learning for this, is the 10-fold cross validation where the data is divided in ten folds and then a model is trained for each fold and gets validated on the rest of the folds. This is the way to evaluate the accuracy of the model and thus to choose the appropriate number of hidden layers and hidden nodes since this is not known beforehand. Usually different topologies are being tested and the one that minimises the error between the predicted and the actual values on the test set is selected. 

\subsection{Topology Defined Neural Network}

The flexibility that Neural Networks provide in designing the topology can be exploited to incorporate knowledge extracted by unsupervised learning performed on the data. Thus, in this work we tried to organise the neurons in the hidden layers based on the knowledge extracted by Factor Analysis and Clustering. The idea behind this was based on the striking resemblance Neural Networks have with Latent Factor Models, like Factor Analysis, and on the assumption that the classes of debtors identified by clustering define different relationships between the response variable and the predictors.

Factor Analysis is a common Latent Factor Model that organises the variables of a dataset into a smaller number of hidden factors that would still contain most of the information from the initial variables. This way neurons in the first hidden layer can be depicted as latent factors that summarise the input. The only difference with Factor Analysis, a widely used Latent Factor Model, is that the relationship between the input variables and the factors is non-linear. This non-linear relationship would also be able to model the linear relationships between the input variables and the neurons identified by Factor Analysis. This idea has been incorporated with the algorithm proposed in \cite{ding10}.

Clustering on the other hand divides the debtors into classes with distinct characteristics. As these classes may model different relationships between the response variables and the explanatory variables this could be introduced in the neural network as an extra hidden layer with as many neurons as the classes. This would create different functions for each class that will be combined in a more complex relationship in order to produce the final modelling. The intuition is something similar to Clusterwise Regression but the combination of different functions for each class is more fuzzy since they are included in a neural network and not hard.

These two ideas form this novel method to use Neural Networks that we named Topology Defined Neural Network (TopDNN). Our aim is to test TopDNN in the socio-economic context but its disciplines can be extended in creating Neural Networks models for any real world application.

\section{Experimental Setup}

The aim of this work is to evaluate the performance of Neural Networks as a regression model that can predict the amount of unsecured debts (\emph{udebt}) a debtor in the CCCS has by using the rest of the variables as predictors. For this reason we compare its performance against different regression models with different characteristics, like Linear Regression, Random Forest Regression. Furthermore we check whether a series of transformations we performed in \cite{lad14} and the classification of debtors we provided in the same work can improve the performance of the regression so that they be incorporated in the final Neural Network we aim to develop.

Since these models try to optimise different criteria and they are internally validated on different measures when they are fitted into data, we needed to test all these models under a common framework. So we use the 10-fold cross validation as the method to compare the different models and we selected RMSE and $R^2$ as the evaluation criteria. 10-fold cross validation is a standard method for evaluating models in Unsupervised Learning and it also allows Neural Networks to avoid data overfitting providing more representative results for their case.

$R^2$ measures the percentage of variance that is explained by the model and it a standardised measure taking values from 0 to 1 with 1 being a perfect fit. The Root Mean Square Error (RMSE) measures the difference between the predicted values from the model and the actual values. It is defined as:

\begin{equation}
RMSE=\sqrt{\frac{\sum_{i=1}^{n}(y_{obs,i}-y_{model},i)^2}{n}}
\end{equation}

\noindent where n is the number of observations, $y_{obs,i}$ is the observed value of the observation i and $y_{model,i}$ is the calculated value of the observation i. The best model will minimise the RMSE.

For model training we use a random sample of 10000 debtors from the CCCS dataset, a subset of dataset that contains no missing values and we already had performed the transformations on and divided in classes \cite{lad14}. All the models are built in R using the \emph{caret} package and for Linear Regression we calculate the weights using the OLS algorithm, for Random Forests we create 500 trees and initialise the number of potential candidates for a node split as m/3 where m equals the number of predictors. For Neural Networks the initial weights are randomly assigned and a hidden layer is chosen. In order to choose the optimal number of hidden nodes, we produce ten neural networks for each case with the number of neurons varying from 1 to 10. 10-fold cross validation is used to evaluate all of them and the one with that minimised RMSE is selected as the best model. We also use both Backpropagation and Resilient Backpropagation for making the appropriate comparisons. All models are built using both the actual data and the transformed and the classification is introduced as an additional categorical variable. For all of the above we had to create four different datasets that all the regression models will be build upon. These necessary datasets in order test the contribution of the transformation and the classification provided by clustering together with the performance of the regression models are summarised in Table III.

\begin{table}[btp]
\caption{Description of Datasets}
\begin{center}
\begin{tabular}{ll}
\hline\noalign{\smallskip}
$ Dataset $ & $ Attributes $ \\
\noalign{\smallskip}
\hline
\noalign{\smallskip}
A  & Original CCCS variables\\  
B  & Transformed Variables \\  
C  & Original CCCS variables and clustering classification \\
D & Transformed variables and clustering classification \\      
\hline
\end{tabular}
\end{center}
\end{table}

Finally we construct a Neural Network based on our intuition to utilise clustering classifications and Factor Analysis for designing the topology and we checked its performance in the same dataset.

\section{Results}
\subsection{Comparison of Models}

From a quick look in Table IV, which presents the performance of the models build on the four datasets with the brackets indicating the optimal number of neurons in the hidden layer found, we can see that Neural Networks and Random Forests clearly outperform Linear Regression on almost datasets with the only exception being the Neural Network model build on the C dataset and was trained with backpropagation. In all the rest of the cases Neural Networks and Random Forests produce smaller RMSE and bigger $R^2$. In addition to this we can identify the beneficial nature of the transformations performed on CCCS attributes since all four different regression models seem to improve their performance when they are built on the transformed data. More specifically, the models built on datasets containing the transformed attributes (B and D) reduce the RMSE and increase $R^2$ when compared with models built on datasets A and C respectively. Especially in the cases of Neural Networks trained with Resilient Backpropagation and the Random Forests regression the improvement in the performance is significantly big reducing the RMSE to around 0.06 for the case of Random Forests and to around 0.05  for the case of Neural Networks trained with Resilient Backpropagation. Similarly $R^2$ was raised to around 0.5 for Random Forests and around 0.6 for the Neural Networks trained with Backpropagation. For the cases of Linear Regression and Neural Networks trained with Backpropagation the improvement was significant but much smaller. 

On the other hand the contribution of the classifications provided from clustering remains less clear. It manages to provide a rather small improvement in the Linear Regression and the Backpropagation Neural Networks but it decreases the performance of Random Forests while in Resilient Backpropagation Neural Networks it is beneficial only when it is combined with the transformed data. This can be seen when you compare models built on datasets C and D that contain the additional categorical variable of classification with the models build on datasets A and B respectively. Interestingly enough the Random Forest regression model build on C has an increased RMSE and a bigger proportion of variance explained at the same time.

Looking at the performance of the models, the best performance was achieved by the Resilient Backpropagation models followed closely by the Random Forests Regression whereas the performance of Backpropagation Neural Networks and Linear Regression remained comparable with the first one being better though. The model that exhibits the minimum RMSE and the bigger $R^2$ is the Resilient Backpropagation Neural Network built on the transformed variables together with the classification of debtors. This verified the argument of \cite{gun10} that Resilient Backpropagation is more suitable for regression purposes. It also strengthens the argument regarding the potential of using Neural Networks in applications of Economics, traditionally dominated by statistical models. Data Mining and Computational Intelligence in a broader sense introduce a holistic approach in order to extract knowledge from that data as it offers a large number of tools to preprocess the data, techniques to explore the relationships with unsupervised learning algorithms like clustering and accurate models to be used for prediction, that when combined in a sophisticated framework, they can build models which achieve impressive results. In our case this was verified not only by the better performance of Neural Networks and Random Forests but also from the beneficial nature of the transformations performed on the data as part of preprocessing the data that improved all the models. Despite the fact that the contribution of the classification of debtors returned from clustering was not beneficial for all the cases tested, it managed to provide a small improvement in most of the cases and especially when it was combined with the transformations in the Neural Networks.

\begin{table}[btp]
  \caption{Results of Regression Models}
	\begin{center}	    
    \begin{tabular}{lll}
 	\hline\noalign{\smallskip}
    \textbf{Dataset} & \textbf{RMSE} & \textbf{Rsquared} \\
	\hline
	\noalign{\smallskip}    
    \textit{Linear Regression} & \textit{} & \textit{} \\
    A     & 0,078 & 0,235 \\
    B     & 0,0731 & 0,328 \\
    C     & 0,0769 & 0,257 \\
    D     & 0,0727 & 0,336 \\
    \textit{Random Forests} & \textit{} & \textit{} \\
    A     & 0,0727 & 0,293 \\
    B     & 0,0592 & 0,572 \\
    C     & 0,0741 & 0,311 \\
    D     & 0,0626 & 0,5 \\
    \textit{Neural Networks Backpropagation} & \textit{} & \textit{} \\
    A (4 Neurons) & 0,0779 & 0,241 \\
    B (2 Neurons) & 0,0672 & 0,445 \\
    C (4 Neurons) & 0,0778 & 0,239 \\
    D (2 Neurons) & 0,0671 & 0,445 \\
    \textit{Neural Networks Resilient Backpropagation} &       &  \\
    A (3 Neurons) & 0,0759 & 0,314 \\
    B (3 Neurons) & 0,0552 & 0,619 \\
    C (2 Neurons) & 0,0764 & 0,26 \\
    D (3 Neurons) & 0,0538 & 0,632 \\
    \hline
    \end{tabular}%
    \end{center}
  \label{Regression Results}%
\end{table}%

Proceeding with the examination the $R^2$ achieved by the models, we notice that the best model has the ability to explain approximately two times the proportion of variance explained by the best Linear Regression model. When these models are compared to the ones found in literature, Linear Regression performance seems to be comparable to the one presented in \cite{kim01} but better than the rest of the models whereas the performance of the Neural Networks trained by Resilient Backpropagation is significantly higher and can only be compared with the Linear Regression model in \cite{livin92} but this was considered not representative enough due to the limited number of observations the model was build upon. In fact, a more realistic number of $R^2$ for this model given in \cite{stone06} was arround 30\% meaning that the performance of the best model found here is still significantly higher than the ones found in literature.

\subsection{Analysing Linear Regression}

The low performance of Linear Regression comparing to the Data Mining methods can be explained easily if we take a careful look at the diagnostics plots of the best linear model in Fig. 2 The plot of the residuals against the fitted values indicates that the error terms are not independent and that their variance is not constant as they are not randomly scattered throughout the 0. Besides this, the normal probability reveals that the error terms are not normally distributed as there is a strong deviation from the line with two big curves in the beginning and the end of the plot. Furthermore in Fig 3., where the partial residuals plot for \emph{Housing Factor} is depicted, we can identify the non-linear relationship it has with the response variable. Partial residuals are utilised instead of normal residuals because in a multiple regression they account for the effect the rest of the independent variables have on this relationship.  These observations come in contrast with almost all the assumptions of linear regression, degrading the quality of the linear model. A series of transformations on the response variable or the explanatory variables, following established techniques like power and log transformations were not able to improve the quality of the model as the $R^2$ remained low and the assumptions were still violated.  

\begin{figure}[btp]
\centerline{\includegraphics[scale=0.5]{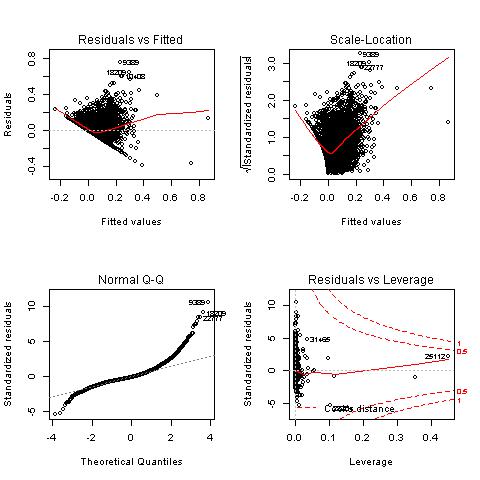}}
\caption{Diagnostic plots of Linear Regression model built on D Dataset} \label{fig 2}
\end{figure}


\begin{figure}[btp]
\centerline{\includegraphics[scale=.6]{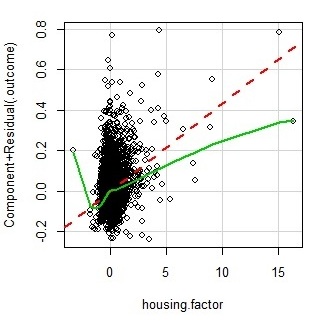}}
\caption{Partial Residuals plot of \emph{Housing Factor}} \label{fig 3}
\end{figure}

\subsection{TopDNN}

Since the benefical nature of the transformed variables is experimentally verified in all cases  we are encouraged to test our novel approach on the dataset B using Resilient Backpropagation. Therefore we begin with performing a Factor Analysis on the attributes of the dataset B. Three was the number of factors that is found to be optimal for summarising the nine attributes of the dataset after examining the scree plot of the eigenvalues and performing a parallel analysis. In the scree plot the eigenvalues of the correlation matrix are plotted in order of descending values. The last substantial drop in the graph indicates the number of factors. In parallel analysis the same eigenvalues are compared to eigenvalues derived from random data. The number of cases they are bigger suggests the number of factors in the model. These methods for determining the number of factors are two of the most popular and effective and they are preferred from others as dictated in \cite{fab99}. Interestingly enough three is also the number of neurons that was found to be optimal for the case of building Neural Networks on C using Resilient Backpropagation indicating the agreement between two different techniques in designing the network topology of a neural network. The three factors and their loadings can be seen in Table V.

\begin{table}[btp]
  \caption{Factor Analysis on transformed variables}
	\begin{center} 
    \begin{tabular}{llll}
    \hline\noalign{\smallskip}
          & \textbf{Factor1} & \textbf{Factor2} & \textbf{Factor3} \\
    \hline
    \noalign{\smallskip}
    x     & 0.298 & 0.487 &  \\
    y     &       &       &  \\
    housingfactor &       & 0.385 & -0.477 \\
    financialfactor1 & 0.280 & 0.574 & 0.766 \\
    financialfactor2 & 0.118 & 0.792 &  \\
    Necessity.Spending & 0.983 &       & 0.167 \\
    Household.Spending & 0.728 & 0.286 & 0.232 \\
    Excessive.Spending & 0.217 &       & 0.128 \\
    Leisure.Spending &       &       &  \\
    \hline
    \end{tabular}%
    \end{center}
  \label{Table 7}%
\end{table}%

Then we train two Neural Networks, one with one hidden layer of three neurons and one with an additional hidden layer of eight neurons representing the classes of debtors, in order to test in a stepwise fashion the two main ideas of our approach. Again we utilise the 10-fold cross validation and RMSE and $ R^2$ as evaluation criteria in order to get comparable results with the rest of the experiments. The results can be seen in Table VI. We can see that designing the network topology according to the knowledge extracted by Factor Analysis and Clustering is beneficial for the performance of the model. We see that the inclusion of the hidden layer of three nodes as dictated by Factor Analysis improves the performance of the model when compared with the Neural Network build on B but has worse performance from the best model of the previous experiments. The additional layer of eight neurons on the other hand achieves the best performance from all the models build here in the work raising the $R^2$ to 0.633 and reducing the RMSE to 0.0528. This verified our intuition that the flexibility Neural Networks offer in designing their topology can be exploited properly in order to include knowledge that stems from the unsupervised learning approaches performed on the data. Thus our model manages to achieve the best performance of all the models indicating the ability of Neural Networks to incorporate in their modelling results from previous steps of the Data Mining process. 

\begin{table}[btp]
  \caption{Results of TopDNN}
	\begin{center}    
    \begin{tabular}{lll}
    \hline\noalign{\smallskip}
          & \textbf{RMSE} & \textbf{Rsquared} \\
    \hline
    \noalign{\smallskip}
    \textit{TopDNN} & \textit{} & \textit{} \\
    NN with factor analysis & 0,055 & 0,616 \\
    NN with factor analysis and clustering & 0,0528 & 0,633 \\
    \hline
    \end{tabular}%
    \end{center}
  \label{Table 8}%
\end{table}%

The plot of the Neural Network build with the TopDNN approach can be seen in Fig. 4. The weights of the edges have been omitted for classification reasons but the lines have modified accordingly to depict the magnitude of the weights with thinner line representing small or negative weights and thicker lines large weights. We can notice that the interpretation of Neural Network is not a trivial task, especially when the network is complicated. That is their main drawback comparing to Linear Regression and Random Forest which have mechanism to assess the variable importance of their models. However tracing the very thick black lines of the plot we can immediately detect the strong influence \emph{FinancialFactor1} has on the final outcome as it influences heavily the first neuron of the first hidden which influenced strongly the sixth neuron of the 2nd hidden layer which belongs to the four neurons of the 2nd layer that affect moderately the final outcome. This relationship between the \emph{FinancialFactor1} and \emph{udebt} cannot be quantified or defined but it can be signified. There are techniques to assess variable importance in Neural Networks, like Sensitivity Analysis that can provide the desired interpretabily that is valuable for the analysis of real world applications but we leave this for the future part of our research.

\begin{figure}[btp]
\centerline{\includegraphics[scale=0.3]{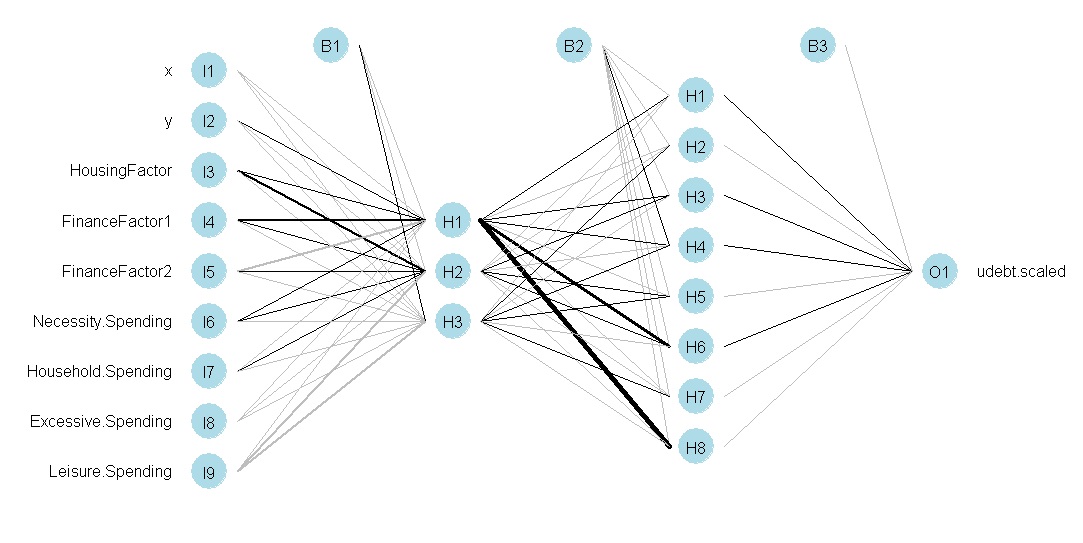}}
\caption{TopDNN with two hidden layers. The first one represents the number of factors and the second one the number of classes of debtors} \label{fig 5}
\end{figure}

\section{Conclusions}
In this work we tried to construct an accurate regression model for the level of debt prediction, a significant task for Consumer Debt Analysis utilising a widely used computational model, Neural Networks. For this reason we compared their performance against Linear Regression and Random Forests. Our results show that Neural Networks clearly outperform Linear Regression. Random Forests achieve comparable performance but their only one parameter does not allow for more improvements. They also proved that all the regression models can benefit from the necessary data transformations and from the Unsupervised Learning approaches on the data, if these are incorporated properly in the Data. Trying the latter we devised a novel method for designing the topology of the Neural Networks utilising information that stems from the Factor Analysis and Clustering performed on the data. TopDNN as our method was named, improved the  performance of the models even more and signified the ability Neural Networks offer in adopting in their design results from previous steps of explanatory research conducted on the dataset. Our work forms a complete Computational Intelligence framework with the pre-processing of data, clustering to uncover important relationships and the regression model that is suitable for the purposes of Consumer Data Analysis. This framework exhibits much better performance than the existing statistical methods that dominate the field of Economics and it highlights a more sophisticated way to model consumer indebtness that it can extend to any real world application.


\subsection{Acknowledgements}
We would like to thank John Gathergood, lecturer in school of Economics of University of Nottingham for providing us the CCCS dataset.

\def\V{\rm vol.~}
\def\N{no.~}
\def\pp{pp.~}
\def\Pot{\it Proc. }
\def\IJCNN{\it International Joint Conference on Neural Networks\rm }
\def\ACC{\it American Control Conference\rm }
\def\SMC{\it IEEE Trans. Systems\rm , \it Man\rm , and \it Cybernetics\rm }

\def\handb{ \it Handbook of Intelligent Control: Neural\rm , \it
    Fuzzy\rm , \it and Adaptive Approaches \rm }


\begin{thebibliography}{22}
\bibitem{ati01}
Atiya, Amir F. ``Bankruptcy prediction for credit risk using neural networks: A survey and new results.'' Neural Networks, IEEE Transactions on 12.4 (2001): 929-935.
\bibitem{kam07}
Bernadette Kamleitner and Erich Kirchler. Consumer credit use: A process model
and literature review. {\it Revue Europeenne de Psychologie Appliquee/European Re-
view of Applied Psychology}, 57(4):267-283, 2007.
\bibitem{kam12}
Bernadette Kamleitner, Erik Hoelzl, and Erich Kirchler. Credit use: Psychological
perspectives on a multifaceted phenomenon. {\it International Journal of Psychology},
47(1):1-27, 2012.
\bibitem{bre01}
Breiman, Leo. ``Random forests'' {\it Machine learning} 45.1 (2001): 5-32.
\bibitem{stone06}
Brice Stone and Rosalinda Vasquez Maury. Indicators of personal financial debt
using a multi-disciplinary behavioral model. {\it Journal of Economic Psychology}, 27
(4):543-556, 2006.
\bibitem{lee09}
De Leeuw, Jan, and Patrick Mair. ``Gifi methods for optimal scaling in R: The package homals.'' {\it Journal of Statistical Software}, forthcoming (2009): 1-30.
\bibitem{ding10}
Ding, Shifei and Jia, Weikuan and Xu, Xinzheng and Zhu, Hong. ``Neural Networks Algorithm Based on Factor Analysis''{\it Advances in Neural Networks} (2010): 319-324
\bibitem{dis09}
Disney R., and Gathergood J.,``Understanding consumer over-indebtedness using counselling sector data: Scoping Study.'', {\it Report to the Department for Business, Innovation and Skills (BIS)}, University of Nottingham, 2009.
\bibitem{fab99}
Fabrigar, Leandre R., et al. "Evaluating the use of exploratory factor analysis in psychological research." Psychological methods 4.3 (1999): 272.
\bibitem{gather12}
Gathergood, John. ``Self-control, financial literacy and consumer over-indebtedness.'' {\it Journal of Economic Psychology} 33.3 (2012): 590-602.
\bibitem{ghos11}
Ghose, Anindya, and Panagiotis G. Ipeirotis. ``Estimating the helpfulness and economic impact of product reviews: Mining text and reviewer characteristics.'' {\it Knowledge and Data Engineering}, IEEE Transactions on 23.10 (2011): 1498-1512.
\bibitem{grom09}
Grömping, Ulrike. ``Variable importance assessment in regression: linear regression versus random forest.'' {\it The American Statistician} 63.4 (2009).
\bibitem{gun10}
Gunther, F., and Fritsch S., ``neuralnet: Training of Neural Networks'',{\it The R Journal},vol 2/1, (2010):30-37
\bibitem{horn89}
Hornik, Kurt, Maxwell Stinchcombe, and Halbert White. ``Multilayer feedforward networks are universal approximators.'' {\it Neural networks} 2.5 (1989): 359-366.
\bibitem{kim01}
Kim, Haejeong, and Sharon A. DeVaney. ``The determinants of outstanding balances among credit card revolvers.'' {\it Financial Counseling and Planning 12.1} (2001): 67-77.
\bibitem{lad14}
Ladas A., Aickelin U., Garibaldi J., Scarpel R., and Ferguson E. ``The Impact of Preprocessing on Clustering socio-economic Data: A Step towards Consumer Debt Analysis'', under Review.
\bibitem{livin92} 
Livingstone, Sonia M., and Peter K. Lunt. ``Predicting personal debt and debt repayment: Psychological, social and economic determinants.'' {\it Journal of Economic Psychology} 13.1 (1992): 111-134.
\bibitem{wang11}
Lili Wang, Wei Lu, and Naresh K Malhotra. Demographics, attitude, personality
and credit card features correlate with credit card debt: A view from china. {\it Journal
of economic psychology}, 32(1):179-193, 2011.
\bibitem{ref94}
Nicholas Refenes, Apostolos, Achileas Zapranis, and Gavin Francis. ``Stock performance modeling using neural networks: a comparative study with regression models.'' {\it Neural Networks} 7.2 (1994): 375-388.
\bibitem{ott11}
Ottaviani, Cristina, and Daniela Vandone. ``Impulsivity and household indebtedness: Evidence from real life.'' {\it Journal of economic psychology 32.5} (2011): 754-761.
\bibitem{seg04}
Segal, Mark R. ``Machine learning benchmarks and random forest regression.'' (2004).
\bibitem{sousa07}
Sousa, S. I. V., et al. ``Multiple linear regression and artificial neural networks based on principal components to predict ozone concentrations.'' {\it Environmental Modelling \& Software} 22.1 (2007): 97-103.

\end{thebibliography}
\end{document}